\documentclass[12pt]{article}
\catcode`\@=11
\def\numberbysection{\@addtoreset{equation}{section}
\def\theequation{\thesection.\arabic{equation}}}
\numberbysection

\def\stackunder#1#2{\mathrel{\mathop{#2}\limits_{#1}}}
\newcommand{\text}[1]{\hbox{#1}}

\def\binom#1#2{{#1 \choose #2}}

\newcommand{\beq}{\begin{equation}}
\newcommand{\eeq}{\end{equation}}
\newcommand{\bea}{\begin{eqnarray*}}
\newcommand{\eea}{\end{eqnarray*}}
\newcommand{\beqa}{\begin{eqnarray}}
\newcommand{\eeqa}{\end{eqnarray}}

\newcommand{\e}{{\rm e}}

\newcommand{\sign}{\mathop{\rm sign}}
\renewcommand{\max}{_{\rm max}}
\newcommand{\frad}[2]{\displaystyle{\displaystyle#1\over\displaystyle#2}}
\renewcommand{\l}{\ell}
\newcommand{\eps}{\varepsilon}
\newcommand{\B}{\mathcal{B}}
\newcommand{\F}{S}

\begin{document}

\centerline{\Large\bf Statistics of the occupation time}
\smallskip
\centerline{\Large\bf for a class of Gaussian Markov processes}

\vspace{1cm}

\centerline{\large
G.~De Smedt$^{a,}$\footnote{desmedt@spht.saclay.cea.fr},
C.~Godr\`eche$^{b,}$\footnote{godreche@spec.saclay.cea.fr},
and J.M.~Luck$^{a,}$\footnote{luck@spht.saclay.cea.fr}}
\vspace{1cm}
\centerline{$^a$Service de Physique Th\'eorique,
CEA Saclay, 91191 Gif-sur-Yvette cedex, France}
\centerline{$^b$Service de Physique de l'\'Etat Condens\'e,
CEA Saclay, 91191 Gif-sur-Yvette cedex, France}

\vspace{.1cm}

\vspace{1cm}

\begin{abstract}
We revisit the work of Dhar and Majumdar [Phys. Rev. E \textbf{59},
6413 (1999)] on the limiting distribution of the temporal mean
$M_{t}=t^{-1}\int_{0}^{t}du\,\sign y_{u}$, for a Gaussian Markovian
process $y_{t}$ depending on a parameter $\alpha $, which can be interpreted
as Brownian motion in the scale of time $t^{\prime }=t^{2\alpha }$.
This quantity, for short the mean `magnetization',
is simply related to the occupation time of the process,
that is the length of time spent on one side of the origin up to time~$t$.
Using the fact that the intervals between sign changes of the process form a
renewal process in the time scale $t'$,
we determine recursively the moments
of the mean magnetization.
We also find an integral equation for
the distribution of $M_{t}$.
This allows a local analysis of this
distribution in the persistence region $(M_t\to\pm1)$,
as well as its asymptotic analysis in the regime where $\alpha$ is large.
We finally put the results thus found in perspective with
those obtained by Dhar and Majumdar by another method, based on a formalism
due to Kac.
\end{abstract}

\vfill
\noindent To be submitted for publication to Journal of Physics A
\hfill T/00/000
\vskip -1pt
\noindent P.A.C.S.: 02.50.Ey, 02.50.Ga, 05.40.+j
\hfill S/00/000
\newpage

\setcounter{footnote}{0}
\section{Introduction}

Consider the stochastic process $y_t$ defined by the Langevin equation
\begin{equation}
\frac{dy_{t}}{dt}=\sqrt{2\alpha }\,t^{\alpha-1/2}\eta _{t},
\label{process}
\end{equation}
where $\alpha $ is a positive parameter, and $\eta _{t}$ is a Gaussian white
noise such that $\left\langle \eta _{t}\right\rangle =0$ and $\left\langle
\eta _{t_{1}}\eta _{t_{2}}\right\rangle =\delta(t_{2}-t_{1})$.
In the new time scale
\[
t^{\prime }=t^{2\alpha},
\]
this process satisfies the usual Langevin equation
for one-dimensional Brownian motion,
\[
\frac{dy_{t^{\prime }}}{dt^{\prime }}=\zeta _{t^{\prime }},
\]
where $\zeta_{t'}$ is still a Gaussian white noise,
with $\left\langle \zeta _{t^{\prime }}\right\rangle =0$
and $\left\langle \zeta _{t_{1}^{\prime }}\zeta
_{t_{2}^{\prime }}\right\rangle =\delta(t_{2}^{\prime }-t_{1}^{\prime })$.
The process defined by (\ref{process}) is a simple example of
subordinated Brownian motion \cite{feller}.
As Brownian motion itself, it is Gaussian, Markovian, and non-stationary.

This process appears in various situations of physical interest. For
instance, it is described in \cite{sebastian} as a Markovian approximation
to fractional Brownian motion. It also appears in ref. \cite{majcrit} for
the special case $\alpha =\frac{1}{4}$, as describing the time evolution of
the total magnetization of a Glauber chain undergoing phase ordering.

Dhar and Majumdar \cite{maj} raised the question of computing the
distribution of the occupation time of this process, that is the length of
time spent by the process on one side of the origin up to time $t$,
\begin{equation}
T_{t}^{\pm }=\int_{0}^{t}du\frac{1\pm \sigma _{u}}{2},
\label{occup}
\end{equation}
where $\sigma_t=\sign y_t$,
or equivalently of
\begin{equation}
M_{t}=\frac{1}{t}\int_{0}^{t}du\,\sigma _{u},
\label{magnet}
\end{equation}
where $M_{t}$, the temporal mean of $\sigma _{t}$, is hereafter referred to
as the `mean magnetization' by analogy with physical situations where
$\sigma _{t}$ represents a spin.
The distribution of the occupation time
bears information on the statistics of persistent events of the process
beyond that contained in the persistence exponent \cite
{dg,baldass,drouffe,dubna,newman,newman2,I}.
This exponent governs the decay $\sim t^{-\theta}$ of the survival probability
of the process,
that is the probability that the process did not cross the origin up to
time $t$.
Actually, for the present case, the
determination of $\theta$ is trite, as shown by a simple reasoning \cite
{maj}: the probability for the Brownian process $y_{t^{\prime }}$ not to
change sign up to time $t^{\prime }$ is known to decay as $(t')^{-1/2}$,
hence for the original process as $t^{-\alpha }$.
This shows that $\theta=\alpha $.

When $\alpha =\frac{1}{2}$, the distribution of the fraction of time spent
on one side of the origin by a random walker, or by Brownian motion, is
given, in the long-time regime, by the arcsine law \cite{levy,feller}.
In contrast, when $\alpha \neq \frac{1}{2}$, the explicit determination of this
distribution, or equivalently of the distribution of $M_{t}$, seems very
difficult.
However, as shown in \cite{maj}, in the long-time regime, the
computation of the asymptotic
moments $\left\langle M_{t}^{k}\right\rangle$ can be
done recursively, using two different methods, yielding the same results.
The first method relies on a formalism due to Kac \cite{kac},
while the second one originates from ref. \cite{dg}.

The method used in ref.~\cite{dg} can be applied to any (smooth) process
for which the intervals of time between sign changes are independent, when
taken on a logarithmic scale, with finite (i.e., non zero) mean $\bar{\l}$.
It eventually leads to a recursive
determination of the moments of $M_t$, as $t\to\infty$
(see equation (\ref{rec_mk}) below).

Dhar and Majumdar make the observation that, since relations~(\ref{rec_mk})
are independent of $\bar{\l}$, they can be applied to 
the determination of the moments of $M_t$ for the process (\ref{process}).
Comparing the resulting expressions of the moments thus obtained 
to those derived by their alternative method shows that this is indeed the case.

However it is not obvious to understand why relations~(\ref{rec_mk})
hold for the (non-smooth) process (\ref{process}),
since the assumptions made in order to derive them do not hold
for such a process.
In particular while, for the class of models with independent time 
intervals on a logarithmic scale,
and finite $\bar{\l}$
(for which the method of ref.~\cite{dg} has been devised),
it is natural to work in a logarithmic scale of time,
since the mean number of sign changes between $0$ and $t$ scales as
$\left\langle N_t \right\rangle\approx(\ln t)/\bar{\l}$,
this is not so in the present case, since $\bar{\l}$
asymptotically vanishes, and the mean number of sign changes scales as
$\left\langle N_t \right\rangle\approx2\,\pi^{-1/2}\,t^\alpha$ \cite{I}.
The validity of relations~(\ref{rec_mk}) for the process (\ref{process})
therefore requires an explanation.

In the present work, we revisit and extend the study done in \cite{maj}.

We first give a new derivation of  the asymptotic expressions of the 
moments $\left\langle M_{t}^{k}\right\rangle$.
We start from the same premises as  in ref. \cite{dg}, then follow another route
--more adapted to the process under study--
because of the difficulties encountered in applying 
step by step the method of ref. \cite{dg} to the present case (sections 2-5).
We then identify the symmetry properties of the distributions of
the random variables that appear in the computations,
and derive a functional integral equation,
the solution of which yields the distribution of $M_{t}$ (section 6).
This approach is first checked on the case $\alpha=\frac{1}{2}$ (section 7).
It is then successively applied to the study of the
local behavior of this distribution in the persistence region,
for general~$\alpha$ (section~8), and to the large-$\alpha$ regime (section 9).

We finally discuss some aspects of ref. \cite{maj}.
We explain why a formal application of the method of ref. \cite{dg} to the present case
is only heuristic, and
give a new interpretation of the results obtained in \cite{maj} with the
method of Kac, under the light of the present work (section 10).

\section{Observables of interest}

Changes of sign of the process $y_{t}$ (or zero crossings) occur at discrete
instants of time $t_{1},t_{2},\dots,t_{n},\dots$,
once the process is suitably regularized at short times.
We assume that the process starts at the origin,
so that $t_0=0$ is also a sign change.
Let $N_{t}$ be the number of sign
changes which occurred between $0$ and $t$, i.e., $N_{t}$ is the random
variable for the largest $n$ for which $t_{n}\le t.$

In the scale $t^{\prime }$, where the process is
(regularized) Brownian motion,
sign changes occur at the instants of time\footnote{%
Hereafter we denote by a prime any temporal variable in this scale.}
$t_{n}^{\prime }=(t_{n})^{2\alpha }$, and $N_{t^{\prime }}\equiv N_{t}$ is
the random variable for the largest $n$ for which $t_{n}^{\prime }\le
t^{\prime }$.
The intervals of time between sign changes are denoted by
$\tau _{n}^{\prime }=t_{n}^{\prime }-t_{n-1}^{\prime }$. These are independent,
identically distributed random variables,
with a probability density function $\rho(\tau')$.
For large values of $\tau'$, $\rho(\tau')$ decays proportionally to
$(\tau')^{-3/2}$.
This behavior is independent of the regularizing procedure,
while its prefactor just reflects the choice of time units.
The density $\rho(\tau')$ is therefore in the
basin of attraction of a L\'{e}vy law of index $\frac{1}{2}$.
We choose units so that we have in Laplace space
\begin{equation}
\stackunder{\tau ^{\prime }}{\mathcal{L}}\rho(\tau ^{\prime })=\hat{\rho}%
(s)=\left\langle \text{e}^{-s\tau ^{\prime }}\right\rangle \stackunder{%
s\rightarrow 0}{\approx }1-\sqrt{s}.
\label{pdf_ro}
\end{equation}
The process formed by the
independent intervals of time $\tau _{1}^{\prime },\tau _{2}^{\prime
},\dots$, is known as a renewal process.
In the original scale $t$, the intervals of time $\tau
_{n}=t_{n}-t_{n-1}$ are \emph{not} independent.

We denote by $t_{N}$ the instant of the last change of sign
of the process before time~$t$. This random variable depends implicitly on time
$t$ through $N_{t}$. In the scale $t^{\prime }$, we have $t_{N}^{\prime
}=(t_{N})^{2\alpha }$.

The occupation times $T_{t}^{+}$ and $T_{t}^{-}$ (see equation~(\ref{occup}))
are the lengths of time spent by the sign process $\sigma_t$ respectively in
the $+$ and $-$ states, up to time $t$, hence $t=T_{t}^{+}+T_{t}^{-}$. They
are simply related to the mean magnetization (\ref{magnet}) by
\[
tM_{t}=T_{t}^{+}-T_{t}^{-}=2T_{t}^{+}-t=t-2T_{t}^{-}.
\]
Assume that $y_{t}>0$ at $t=0^{+}$, i.e., $\sigma _{t=0}=+1$.
Then
\[
tM_{t}=\left\{
\begin{array}{ll}
-(t-t_{N})+(t_{N}-t_{N-1})-\cdots & \text{if }N_{t}=2k+1\text{ (i.e., }\sigma
_{t}=-1),\\
(t-t_{N})-(t_{N}-t_{N-1})+\cdots & \text{if }N_{t}=2k\text{ (i.e., }\sigma
_{t}=+1).
\end{array}
\right.
\]
The converse holds if $\sigma _{t=0}=-1$.
Hence we have, with equal probabilities,
\begin{equation}
M_{t}=\pm(1-2\xi _{t}), \label{eqM}
\end{equation}
where
\[
\xi _{t}=\frac{1}{t}(t_{N}-t_{N-1}+\cdots)
\]
is the fraction of time spent in the state $+$ if $\sigma _{t}=-1$,
and conversely.
The last formula can be rewritten as
\begin{equation}
\xi _{t}=\frac{t_{N}}{t}X_{N}, \label{eqxi}
\end{equation}
where the $X_{N}$ obey the recursion
\begin{equation}
X_{N}=1-\frac{t_{N-1}}{t_{N}}X_{N-1},\label{kesten}
\end{equation}
with $X_{1}=1$. Both random variables $X_{N}$ and $t_{N-1}/t_{N}$ depend
implicitly on time $t$ through $N_{t}$.

For instance, if $\sigma _{t=0}=+1$ and $N_{t}=4$, then
\begin{eqnarray*}
M_{t} &=&\frac{1}{t}\Big(
(t-t_{4})-(t_{4}-t_{3})+(t_{3}-t_{2})-(t_{2}-t_{1})+t_{1}\Big) \\
&=&1-2\xi _{t},
\end{eqnarray*}
where $\xi _{t}=(t_{4}-t_{3}+t_{2}-t_{1})/t=t_{4}X_{4}/t$, with
\[
X_{4}=1-\frac{t_{3}}{t_{4}}\left( 1-\frac{t_{2}}{t_{3}}\left( 1-\frac{t_{1}}{%
t_{2}}\right) \right).
\]

\section{Methods of solution}

Equations (\ref{eqM}), (\ref{eqxi}) and (\ref{kesten}) contain in essence
the solution to the problem posed, namely the determination of the limiting
distribution of the mean magnetization $M_{t}$ for $t\rightarrow \infty $.
Unfortunately, no explicit solution can be attained in general.

However, from (\ref{eqM})-(\ref{kesten}), one can obtain recursively the
moments of $M_{t}$, in the long-time limit.
This can be done either along the lines of ref.~\cite{dg}, as done in \cite{maj},
or by the method of the present work.
In this section, we explain the
difficulty encountered when applying the method of ref. \cite{dg} to the
process (\ref{process}), in order to justify the more lengthy path we have
adopted for the derivation of the moments.
We shall come back to the comparison between the two methods in section 10.

\subsection{General framework}

Assume that, in the long-time regime, the dimensionless random variables
$t_{N}/t$, $t_{N-1}/t_{N}$, $X_{N}$, $\xi _{t}$, and $M_{t}$
possess a limiting joint distribution.
Define
\begin{eqnarray*}
&&H=\lim_{t\rightarrow \infty }\,\frac{t_{N}}{t},\qquad F=\lim_{t\rightarrow
\infty }\,\frac{t_{N-1}}{t_{N}},\, \\
&&X=\lim_{t\rightarrow \infty }X_{N},\qquad \xi =\lim_{t\rightarrow \infty
}\,\xi _{t},\qquad M=\lim_{t\rightarrow \infty }\,M_{t}.
\end{eqnarray*}
Then the equations to be solved are:
\begin{eqnarray}
X &=&1-FX\label{kestenbis},\\
\xi &=&HX\label{eqxibis},\\
M &=&\pm(1-2\xi ).\label{eqMbis}
\end{eqnarray}
These equalities hold in distribution,
and the random variables entering them are not independent a priori.
Equation (\ref{kestenbis}) is to be understood as the fixed-point equation
corresponding to the recursion (\ref{kesten}),
while (\ref{eqxibis}) and (\ref{eqMbis})
respectively correspond to (\ref{eqxi}) and (\ref{eqM}).

Assume that the distribution of the random variable $F$ is given,
and that $F$ is independent of $X$.
Even so, solving (\ref{kestenbis}) is difficult in general
\cite{kesten,kest2,kest3,kest4}.
However, obtaining the moments of $X$ recursively is easier.
If furthermore $H$ and $X$ are independent and the moments of $H$ are known,
then (\ref{eqxibis}) and (\ref{eqMbis}) determine the moments of $M$.

\subsection{The diffusion equation: a reminder}

\label{reminder}

Such a situation precisely arises in the example treated in \cite{dg}: the
process $y_{t}$ is the diffusing field at a fixed point of space, evolving
from random initial conditions, and the so-called independent-interval
approximation is used \cite{majdiffu,derrdiffu}.
In the long-time regime, the process is stationary in
the logarithmic scale of time $T=\ln t$.
As a consequence, the autocorrelation function of the sign process, 
$A(|\Delta T|)=\left\langle \sigma _{T}\sigma _{T+\Delta T}\right\rangle $,
only depends on the difference of logarithmic times \cite{majdiffu,derrdiffu}.

Consider the intervals of time $\l_{N}$ between successive sign changes of
the process in the logarithmic scale of time, $\l_{N}=T_{N}-T_{N-1}$, or
\begin{equation}
\text{e}^{-\l_{N}}=\frac{t_{N-1}}{t_{N}}. \label{deflN}
\end{equation}
The independent-interval approximation consists in considering the intervals
$\l_{N}$ as independent, thus defining a renewal process.
The distribution of the random variable $\l_{N}$
can then be derived, in Laplace space, from the knowledge of the
correlation function $A(|\Delta T|)$.
This distribution is found to be independent of time,
because the process is stationary in logarithmic time.
Its average, $\left\langle \l_{N}\right\rangle =\bar{\l}$,
is some time-independent positive number.
Explicitly,
\begin{equation}
\hat{f}_{\l_{N}}(s)=\left\langle \text{e}^{-s\l_{N}}\right\rangle
=\left\langle F^{s}\right\rangle =\frac{1-\bar{\l}\,g(s)}{1+\bar{\l}\,g(s)},
\label{fls}
\end{equation}
with
\begin{equation}
g(s)=\frac{s}{2}\Big(1-s\hat{A}(s)\Big), \label{gs}
\end{equation}
where $\hat{A}(s)$ is the Laplace transform of $A(T)$.
In particular, the moments
\begin{equation}
f_{k}=\left\langle F^{k}\right\rangle =\left\langle \left( \frac{t_{N-1}}{%
t_{N}}\right) ^{k}\right\rangle =\left\langle \text{e}^{-k\,\l_{N}}\right%
\rangle =\hat{f}_{\l_{N}}(k) \label{def_fk}
\end{equation}
are independent of time. Thus from (\ref{kestenbis}) the moments of $X$ are
determined recursively in terms of the $f_{k}$ (see (\ref{rec_xk})).

In the long-time regime, the distribution of the backward recurrence time
of the process in the logarithmic scale, $\lambda =T-T_{N}$,
is also independent of time.
This logarithmic recurrence time is related to the random variable $H$ by
\[
\text{e}^{-\lambda }=\frac{t_{N}}{t}=H.
\]
Its distribution in Laplace space reads
\begin{equation}
\hat{f}_{\lambda }(s)=\left\langle \text{e}^{-s\lambda}\right\rangle
=\left\langle H^{s}\right\rangle =\frac{2\,g(s)}{s\left( 1+\bar{\l}%
\,g(s)\right) }. \label{flambdas}
\end{equation}

The random variables $X$ and $H$ (or $\lambda $) are independent.
Hence (\ref{eqxibis}) and (\ref{eqMbis}) determine the moments of $M$.
A remarkable fact is that the moments thus obtained, which are functions of
the $f_{k}$ and of $\bar{\l}$,
become independent of $\bar{\l}$ when the~$f_{k}$ are expressed
in terms of the $\hat{A}(k)\equiv \hat{A}_{k}$,
using equations (\ref{fls}),
(\ref{gs}) and (\ref{def_fk}). 
Thus~\cite{dg}
\begin{eqnarray}
\left\langle M^{2}\right\rangle &=&\hat{A}_{1},\nonumber \\
\left\langle M^{4}\right\rangle &=&1-\frac{\left( 1-3\hat{A}_{1}+4\hat{A}%
_{2}\right) \left( 1-3\hat{A}_{3}\right) }{1-2\hat{A}_{2}},
\label{rec_mk}
\end{eqnarray}
and so on.

More generally, the method used in \cite{dg} can be applied to any process
for which the intervals of time between sign changes are independent, when
taken on a logarithmic scale.
It eventually leads to a recursive
determination of the moments of $M$, resulting in~(\ref{rec_mk}).

\subsection{The case of the process (\ref{process})}

The situation for the process (\ref{process}) is more difficult because,
in the limit $t\rightarrow\infty$, $t_{N-1}/t_{N}\rightarrow F=1$.
Hence (\ref{kestenbis}) no longer determines $X$, and furthermore 
$\left\langle \l_{N}\right\rangle \rightarrow 0$.
Now, for the class of models with independent time 
intervals $\l_{N}$ on a logarithmic scale,
and finite (i.e., non-zero) $\left\langle \l_{N}\right\rangle=\bar{\l}$,
for which the method of ref.~\cite{dg}, sketched above, has been devised,
the mean number of sign changes between $0$ and $t$ scales as
$\left\langle N_t \right\rangle\approx(\ln t)/\bar{\l}$.
So, in contrast, in the present case there is no obvious reason
to work with this logarithmic scale of time, since $\left\langle \l_{N}\right\rangle$
asymptotically vanishes, and the mean number of sign changes scales as
$\left\langle N_t \right\rangle\approx2\,\pi^{-1/2}\,t^\alpha$ \cite{I}.

On the other hand, if time is kept finite, then the time intervals 
$\l_{N}$ are not independent
and the process (\ref{process}) is not stationary, precluding again the application of the
method of ref. \cite{dg}.

A way out is to formally apply this method to the process (\ref{process}),
without paying attention to the difficulties mentioned above,
and taking advantage of the fact
that the moments $\left\langle M^{k}\right\rangle $,
given by (\ref{rec_mk}), are independent on $\bar{\l}$,
and therefore (hopefully) insensitive to the fact that $\left\langle
\l_{N}\right\rangle \rightarrow 0$.
This approach, which is the one followed by Dhar and Majumdar \cite{maj},
is however only heuristic, as further discussed in section \ref{revisit}.

Our approach relies instead on the fact that the time intervals $\tau'_n$
between two sign changes of the process (\ref{process}) form a renewal process
(the $\tau'_n$ are independent, identically distributed random variables
with density $\rho(\tau')$,
given by (\ref{pdf_ro}) for large $\tau'$).
This is a fundamental
property of the process (\ref{process}), and in particular of Brownian
motion if $\alpha =\frac{1}{2}$.

This property allows us to determine the limiting distribution
$f_{H}$ of $H$ when $t\rightarrow \infty $, and the moments $f_{k,t}$
of the random variable $t_{N-1}/t_{N}$, which are now time dependent.
We also find the explicit time-dependent expression of
$\left\langle\l_{N}\right\rangle $.
Using the original equation (\ref{kesten}), instead of
(\ref{kestenbis}), and equations (\ref{eqxibis}) and (\ref{eqMbis}), we
eventually recover the expressions~(\ref{rec_mk}) of the
$\left\langle M^{k}\right\rangle $, thus extending their range of applicability.

We then establish
an integral equation for $f_{X}$, and study its consequences.

\section{Distribution of $t_{N}/t$}

Using the independence of the $\tau _{1}^{\prime },\tau _{2}^{\prime
},\dots$, we first determine the distribution of the random variable $%
t_{N}^{\prime },$ from which we then deduce that of $t_{N}$. The method used
below is borrowed from ref. \cite{I}, where a thorough study of the
statistics of the occupation time of renewal processes can be found.

We denote by $f_{t_{N}^{\prime },N}$ the joint probability density of the
random variables $t_{N}^{\prime }$ and~$N_{t}$. It reads
\begin{eqnarray*}
f_{t_{N}^{\prime },N}(t^{\prime };y,n) &=&\frac{d}{dy}\mathcal{P}\left(
t_{N}^{\prime }<y,N_{t^{\prime }}=n\right)\\
&=&\left\langle \delta(y-t_{N}^{\prime })\,I(t_{n}^{\prime }<t^{\prime
}<t_{n+1}^{\prime })\right\rangle,
\end{eqnarray*}
where $I(t_{n}^{\prime }<t^{\prime }<t_{n+1}^{\prime })=1$ if the event
inside the parenthesis occurs, and $0$ if not. The brackets denote the
average over $\tau _{1}^{\prime },\tau _{2}^{\prime },\dots$ Summing over $%
n$ gives the distribution of $t_{N}^{\prime }$,
\[
f_{t_{N}^{\prime }}(t^{\prime };y)=\sum_{n=0}^{\infty }f_{t_{N}^{\prime
},N}(t^{\prime };y,n)=\left\langle \delta(y-t_{N}^{\prime })\right\rangle.
\]

In Laplace space, where $s$ is conjugate to $t^{\prime }$ and $u$ to $y$,
\begin{eqnarray}
\stackunder{t^{\prime },y}{\mathcal{L}}\,f_{t_{N}^{\prime },N}(t^{\prime
};y,n) &=&\hat{f}_{t_{N}^{\prime },N}(s;u,n)=\left\langle \text{e}%
^{-ut_{n}^{\prime }}\int_{t_{n}^{\prime }}^{t_{n+1}^{\prime }}dt^{\prime }%
\text{e}^{-st^{\prime }}\right\rangle \nonumber \\
&=&\left\langle \text{e}^{-ut_{n}^{\prime }}\text{e}^{-st_{n}^{\prime }}%
\frac{1-\text{e}^{-s\tau _{n+1}^{\prime }}}{s}\right\rangle \nonumber \\
&=&\hat{\rho}(s+u)^{n}\frac{1-\hat{\rho}(s)}{s}\qquad(n\geq 0).\label{ftn}
\end{eqnarray}
Note that setting $u=0$ in (\ref{ftn}) gives the distribution of $%
N_{t^{\prime }}$. We finally obtain
\begin{eqnarray*}
\stackunder{t^{\prime },y}{\mathcal{L}}\,f_{t_{N}^{\prime }}(t^{\prime };y)
&=&\stackunder{t^{\prime }}{\mathcal{L}}\left\langle \text{e}%
^{-ut_{N}^{\prime }}\right\rangle =\hat{f}_{t_{N}^{\prime }}(s;u) \\
&=&\sum_{n=0}^{\infty }\hat{f}_{t_{N}^{\prime },N}(s;u,n)=\frac{1}{1-\hat{%
\rho}(s+u)}\frac{1-\hat{\rho}(s)}{s}.
\end{eqnarray*}
In the long-time regime, i.e., for $s$ and $u$ simultaneously small,
we get the scaling form
\[
\hat{f}_{t_{N}^{\prime }}(s;u)\approx \frac{1}{\sqrt{s(s+u)}},
\]
which yields
\[
f_{t_{N}^{\prime }}(t^{\prime };y)\stackunder{t^{\prime }\rightarrow \infty
}{\approx }\frac{1}{\pi\sqrt{y(t^{\prime }-y)}}.
\]
As a consequence, the random variable $H^{\prime }=\lim_{t^{\prime
}\rightarrow \infty }\,t^{\prime -1\,}t_{N}^{\prime }$ possesses the
limiting distribution
\begin{equation}
f_{H^{\prime }}(x)=\frac{1}{\pi\sqrt{x(1-x)}},
\label{arcsin}
\end{equation}
which is the arcsine law on $[0,1]$.

Using the equality $t_{N}/t$ $=\left( t_{N}^{\prime }/t^{\prime }\right)
^{1/2\alpha }$, this last result yields immediately the distribution of
\[
H=\lim_{t\rightarrow \infty }\,t_N/t=(H')^{1/2\alpha },
\]
which reads
\begin{equation}
f_{H}(x)=\frac{2\alpha x^{\alpha-1}}{\pi\sqrt{1-x^{2\alpha}}}
=\frac{2\alpha}{\pi x\sqrt{x^{-2\alpha}-1}}.
\label{fH}
\end{equation}
This is the main result of this section. Let us define
\begin{equation}
h(s,\alpha )=\left\langle H^{s}\right\rangle =\frac{1}{\pi }B\left( \frac{s}{%
2\alpha }+\frac{1}{2},\frac{1}{2}\right)
=\frac{1}{\sqrt{\pi}}\frac{\Gamma\left(\frad{s}{2\alpha}+\frad{1}{2}\right)}
{\Gamma\left(\frad{s}{2\alpha}+1\right)},
\label{hsa}
\end{equation}
where $B(a,b)=\Gamma(a)\Gamma(b)/\Gamma(a+b)$ is the beta function. For
integer values of $s$, (\ref{hsa}) gives the moments of $f_{H}$, denoted by
\[
h_{k}^{(\alpha )}=h(k,\alpha )=\left\langle H^{k}\right\rangle.
\]

In the particular case $\alpha =\frac{1}{2}$, corresponding to Brownian motion,
the distribution of $H\equiv H'$
is the arcsine law (\ref{arcsin}), with moments
\begin{equation}
h_{k}^{(1/2)}=\frac{1}{\pi}B\left(k+\frac12,\frac12\right )
=\frac{(2k-1)!!}{2^{k}\,k!}=\frac{(2k)!}{2^{2k}(k!)^2}.
\label{mom_arcsin}
\end{equation}

\section{Determination of the moments}

In order to obtain recursion relations for the moments of the random
variable $X$, we proceed in two steps.
We first compute the moments of
$t_{N-1}^{\prime }/t_{N}^{\prime }$, from which we deduce those of
$t_{N-1}/t_{N}$.
The recursion relations for the
$\left\langle X^{k}\right\rangle $ then emerge from (\ref{kesten}).
Equations (\ref{eqxibis}) and (\ref{eqMbis})
finally determine the moments of $M$.

\subsection{Moments of $t_{N-1}/t_{N}$}

We first determine the probability density function of the joint variables $%
t_{N-1}^{\prime }$ and $t_{N}^{\prime }$. In Laplace space, we have
\begin{eqnarray*}
\hat{f}_{t_{N-1}^{\prime },t_{N}^{\prime },N}(s;u,v,n) &=&\left\langle \text{%
e}^{-ut_{N-1}^{\prime }}\text{e}^{-vt_{N}^{\prime }}\int_{t_{n}^{\prime
}}^{t_{n+1}^{\prime }}dt^{\prime }\text{e}^{-st^{\prime }}\right\rangle \\
&=&\left\{
\begin{array}{ll}
\hat{\rho}(s+u+v)^{n-1}\hat{\rho}(s+v)\frad{1-\hat{\rho}(s)}{s}&(n\geq
1),\\
\frad{1-\hat{\rho}(s)}{s}&(n=0).
\end{array}
\right.
\end{eqnarray*}
Summing over $n$ gives
\begin{eqnarray}
\stackunder{t^{\prime }}{\mathcal{L}}\left\langle \text{e}%
^{-ut_{N-1}^{\prime }}\text{e}^{-vt_{N}^{\prime }}\right\rangle &=&\hat{f}%
_{t_{N-1}^{\prime },t_{N}^{\prime }}(s;u,v)=\sum_{n=0}^{\infty }\hat{f}%
_{t_{N-1}^{\prime },t_{N}^{\prime },N}(s;u,v,n) \nonumber \\
&=&\frac{1-\hat{\rho}(s)}{s}\left( 1+\frac{\hat{\rho}(s+v)}{1-\hat{\rho}%
(s+u+v)}\right), \label{ftt}
\end{eqnarray}
so that in particular
$\hat{f}_{t_{N-1}^{\prime },t_{N}^{\prime }}(s;u=0,v=0)=1/s$.

The first moment of the random variable $t_{N-1}^{\prime }/t_{N}^{\prime }$
is obtained by considering
\begin{eqnarray*}
\stackunder{t^{\prime }}{\mathcal{L}}\left\langle \frac{t_{N-1}^{\prime }}{%
t_{N}^{\prime }}\right\rangle &=&\int_{0}^{\infty }dv\,\left( -\frac{d}{du}%
\right) _{u=0}\stackunder{t^{\prime }}{\mathcal{L}}%
\left\langle \text{e}^{-ut_{N-1}^{\prime }}\text{e}^{-vt_{N}^{\prime
}}\right\rangle \nonumber \\
&=&\frac{\hat{\rho}(s)}{s}+\frac{1-\hat{\rho}(s)}{s}\ln(1-\hat{\rho}(s))
\nonumber \\
&\stackunder{s\rightarrow 0}{\approx}&\frac{1}{s}+\frac{\ln s}{2\sqrt{s}},
\end{eqnarray*}
which leads to
\begin{equation}
\left\langle \frac{t_{N-1}^{\prime }}{t_{N}^{\prime }}\right\rangle
\stackunder{t^{\prime }\rightarrow \infty }{\approx }1-\frac{\ln t^{\prime }%
}{2\sqrt{\pi t^{\prime }}},\label{first}
\end{equation}
omitting the finite parts of the logarithms.

This computation generalizes to higher-order moments, using the asymptotic
form~(\ref{pdf_ro}) in (\ref{ftt}).
We have
\begin{eqnarray*}
\stackunder{t^{\prime }}{\mathcal{L}}\left\langle \left( \frac{%
t_{N-1}^{\prime }}{t_{N}^{\prime }}\right) ^{k}\right\rangle &=&\left(
\int_{0}^{\infty }dv\,\right) ^{k}\left( -\frac{d}{du}\right)_{u=0}^{k}
\stackunder{t^{\prime }}{\mathcal{L}}\left\langle \text{e}%
^{-ut_{N-1}^{\prime }}\text{e}^{-vt_{N}^{\prime }}\right\rangle \\
&\stackunder{s\rightarrow 0}{\approx}&\frac{1}{s}+kh_{k}^{(1/2)}
\,\frac{\ln s}{\sqrt{s}},
\end{eqnarray*}
which leads to
\begin{equation}
\left\langle \left( \frac{t_{N-1}^{\prime }}{t_{N}^{\prime }}\right)
^{k}\right\rangle \stackunder{t^{\prime }\rightarrow \infty }{\approx }%
1-kh_{k}^{(1/2)}\,\frac{\ln t^{\prime }}{\sqrt{\pi t^{\prime }}},
\label{jmom}
\end{equation}
where $h_{k}^{(1/2)}$ is given by equation (\ref{mom_arcsin}).
In particular, since $h_{1}^{(1/2)}=\frac12$, (\ref{first}) is recovered.

The result~(\ref{jmom}) can be extended to non-integer values of $k$.
We have thus (see~(\ref{deflN}))
\[
\left\langle \l_{N}^{\prime }\right\rangle
=-\left\langle \ln \frac{t_{N-1}^{\prime }}{t_{N}^{\prime }}\right\rangle
=\lim_{k\rightarrow 0}\left\langle
\frac{1-\left(\frac{t_{N-1}^{\prime}}{t_{N}^{\prime}}\right)^k}{k}\right\rangle
\stackunder{t^{\prime }\rightarrow \infty }{\approx }
\frac{\ln t^{\prime }}{\sqrt{\pi t^{\prime }}},
\]
as $\lim_{k\rightarrow0}\,h_{k}^{(1/2)}=1$.
Equation~(\ref{jmom}) can thus be rewritten as
\begin{equation}
\left\langle \left( \frac{t_{N-1}^{\prime }}{t_{N}^{\prime }}\right)
^{k}\right\rangle \stackunder{t^{\prime }\rightarrow \infty }{\approx }%
1-kh_{k}^{\left( 1/2\right) }\,\left\langle \l_{N}^{\prime }\right\rangle.
\label{H'k}
\end{equation}
As announced above, when $t\rightarrow \infty $, the random variable
$t_{N-1}^{\prime }/t_{N}^{\prime }$ converges to 1, in distribution.

The moments $f_{k,t}$ of $t_{N-1}/t_{N}$ are obtained from (\ref{H'k}) as
\[
f_{k,t}=\left\langle \left( \frac{t_{N-1}}{t_{N}}\right) ^{k}\right\rangle
=\left\langle \left( \frac{t_{N-1}^{\prime }}{t_{N}^{\prime }}\right)
^{k/2\alpha }\right\rangle \stackunder{t^{\prime }\rightarrow \infty }{%
\approx }1-\frac{k}{2\alpha }h_{k}^{(\alpha )}\,\left\langle \l_{N}^{\prime
}\right\rangle,
\]
because $h(k/2\alpha ,\frac{1}{2})=h(k,\alpha )\equiv h_{k}^{(\alpha )}$.
On the other hand,
\[
\bar{\l}_t=\left\langle \l_{N}\right\rangle
=-\left\langle \ln \frac{t_{N-1}}{t_{N}} \right\rangle
=\frac{1}{2\alpha }\left\langle \l_{N}^{\prime }\right\rangle
\stackunder{t\rightarrow \infty }{\approx }
\frac{\ln t}{\sqrt\pi\,t^\alpha},
\]
hence finally
\begin{equation}
f_{k,t}\stackunder{t\rightarrow \infty }{\approx }1-k\,h_{k}^{(\alpha )}\,%
\bar{\l}_{t}.
\label{fkt}
\end{equation}

\subsection{Moments of $X$}

From the recursion relation (\ref{kesten}), we have
\begin{equation}
\left\langle X_{N}^{k}\right\rangle =\left\langle \left( 1-\frac{t_{N-1}}{%
t_{N}}X_{N-1}\right) ^{k}\right\rangle.
\label{jdecou}
\end{equation}
In the long-time regime, there is an asymptotic decoupling of the variables
$X_{N-1}$ and $t_{N-1}/t_{N}$, so that it is legitimate to take
$X_{N}\rightarrow X$, while keeping the leading
time dependence of $f_{k,t}$, given by (\ref{fkt}).
This procedure can be justified along the lines of ref. \cite{I}.
Consider first the simple situation $\alpha=\frac12$.
The difference between unity and $t_{N-1}/t_N$,
which gives rise to the result (\ref{fkt}),
is proportional to the interval $\tau_N=t_N-t_{N-1}$.
This quantity has been shown in \cite{I} to be,
asymptotically for large~$t$, independent of $t_N$,
and distributed according to the a priori law $\rho(\tau)$.
A similar decoupling asymptotically takes place for generic $\alpha$.

Denoting the moments $\left\langle X^{k}\right\rangle $ by $x_{k}$,
we obtain
\begin{equation}
x_k=\mathcal{B}(f_{k,t}\,x_{k}),
\label{binom}
\end{equation}
with $x_{0}=f_{0,t}=1$, and where
we have introduced the notation $\mathcal{B}$ for
the linear binomial operator
\begin{equation}
\mathcal{B}(x_k)
=\sum_{j=0}^{k}\binom{k}{j}\left( -1\right) ^{j}\,x_{j}.
\label{jb}
\end{equation}

As shown in the Appendix, (\ref{binom}) implies the following recursion
relations, according to the parity of $k$,
\begin{eqnarray}
x_{k}(1+f_{k,t}) &=&\mathcal{B}(x_{k}(1+f_{k,t}))\qquad {\hskip 10pt}
(k\text{ odd}),
\label{odd} \\
x_{k}(1-f_{k,t}) &=&-\mathcal{B}(x_{k}(1-f_{k,t}))\qquad(k\text{ even}).
\label{even}
\end{eqnarray}
Using the expression (\ref{fkt}) of $f_{k,t}$, we obtain, in the limit
$t\rightarrow \infty $, where $\bar{\l}_{t}\rightarrow 0$,
\begin{eqnarray}
x_{k} &=&\mathcal{B}(x_{k})\qquad{\hskip 37pt}(k\text{ odd}),\label{recur1} \\
k\,h_{k}^{(\alpha )}x_{k} &=&-\mathcal{B}(kh_{k}^{(\alpha )}\,x_{k})
\qquad(k\text{ even}).
\label{recur2}
\end{eqnarray}
These relations, which can be rewritten as
\begin{eqnarray}
x_{k} &=&\frac{1}{2}\sum_{j=0}^{k-1}\binom{k}{j}\left( -1\right)
^{j}\,x_{j}\qquad {\hskip 64pt}(k\text{ odd}),\label{recur1bis}\\
x_{k} &=&-\frac{1}{2k\,h_{k}^{(\alpha )}}\sum_{j=0}^{k-1}\binom{k}{j}\left(
-1\right) ^{j}\,j\,h_{j}^{(\alpha )}\,x_{j}\qquad(k\text{ even}),
\label{recur2bis}
\end{eqnarray}
determine the stationary values of the $x_{k}$ recursively.
We thus obtain
\[
\begin{array}{l}
x_{1}=\frad{1}{2},\qquad
x_{2}=\frad{h_{1}^{(\alpha)}}{4h_{2}^{(\alpha)}},\qquad
x_{3}=-\frad{1}{4}+\frad{3h_{1}^{(\alpha )}}{8h_{2}^{(\alpha )}},\\ \\
x_{4}=-\frad{h_{1}^{(\alpha)}+3h_{3}^{(\alpha)}}{8h_{4}^{(\alpha )}}
+\frad{9h_{1}^{(\alpha)}h_{3}^{(\alpha)}}{16h_{2}^{(\alpha)}h_{4}^{(\alpha)}},
\qquad\dots
\end{array}
\]

\subsection{Moments of $M$}

For reasons similar to those exposed below equation~(\ref{jdecou}),
the random variables $H$ and $X$ (defined in the limit $t\to\infty$)
are independent.
Thus, by (\ref{eqxibis}) we have
\begin{equation}
\left\langle \xi ^{k}\right\rangle =h_{k}^{(\alpha )}\,x_{k},
\label{jdede}
\end{equation}
which, together with equation~(\ref{kestenbis}), leads to a determination of
the even moments of the mean magnetization $M$ in terms of the $x_k$:
\begin{equation}
\left\langle M^{k}\right\rangle =\left\langle(1-2\xi )^{k}\right\rangle
=\mathcal{B}(2^k\,h_k^{(\alpha)}\,x_k) \qquad(k\text{ even}).
\label{jme}
\end{equation}
We thus finally obtain
\begin{eqnarray}
\left\langle M^{2}\right\rangle &=&1-h_{1}^{(\alpha )}, \nonumber \\
\left\langle M^{4}\right\rangle &=&1+2h_{3}^{(\alpha )}
-\frac{3h_{1}^{(\alpha )} h_{3}^{(\alpha )}}{h_{2}^{(\alpha )}},\nonumber\\
\left\langle M^{6}\right\rangle
&=&1-5h_1^{(\alpha)}-10h_3^{(\alpha)}-16h_5^{(\alpha)}
+\frac{h_1^{(\alpha)}(15h_3^{(\alpha)}+20h_5^{(\alpha)})}{h_2^{(\alpha)}}
\nonumber\\
&+&\frac{(10h_1^{(\alpha)}+30h_3^{(\alpha)})h_5^{(\alpha)}}{h_4^{(\alpha)}}
-\frac{45h_1^{(\alpha)}h_3^{(\alpha)}h_5^{(\alpha)}}
{h_2^{(\alpha)}h_4^{(\alpha)}},\
\label{rec_mkbis}
\end{eqnarray}
and so on.

For instance, if $\alpha =\frac{1}{2}$, corresponding to Brownian motion,
the successive even moments of $M$
are equal to $\frac{1}{2}$, $\frac{3}{8}$, $\frac{5}{16}$,
$\frac{35}{128},\dots$, i.e.,
\[
\left\langle M^{2j}\right\rangle=\frac{(2j)!}{2^{2j}(j!)^{2}}=h_{j}^{(1/2)},
\]
which are the even moments of the arcsine law on $[-1,1]$
(see (\ref{fM}) below).

\section{An integral equation for the determination of $f_{M}$}
\label{jsecgen}

The recursion relation (\ref{recur1}) expresses a symmetry property of
the distribution $f_{X}$:
\begin{equation}
f_{X}(x)=f_{X}(1-x) \label{fx_sym}
\end{equation}
(see Appendix).
This is also obvious from (\ref{kestenbis}),
since formally $F=1$ in the present case.

The recursion relation (\ref{recur2}), which can be rewritten as
\begin{equation}
k\left\langle \xi ^{k}\right\rangle =-\mathcal{B}(k\left\langle \xi
^{k}\right\rangle )\qquad(k\text{ even}),
\label{recur2'}
\end{equation}
expresses a symmetry property of the distribution $f_{\xi}$, as we now show.
First, it is easy to prove that
\[
\mathcal{B}(k\left\langle \xi ^{k}\right\rangle )=-k\left\langle \xi(1-\xi
)^{k-1}\right\rangle.
\]
Therefore (\ref{recur2'}) yields
\[
\left\langle \xi ^{k}\right\rangle =\left\langle \xi(1-\xi
)^{k-1}\right\rangle \qquad(k\text{ even}),
\]
which is equivalent
to the following symmetry property
\begin{equation}
\xi f_{\xi }(\xi )=(1-\xi )f_{\xi }(1-\xi ),
\label{fxi_sym}
\end{equation}
or
\begin{equation}
\phi(\xi)=\phi(1-\xi),
\label{jphisym}
\end{equation}
introducing the function
\begin{equation}
\phi(\xi)=\xi f_\xi(\xi).
\label{jphidef}
\end{equation}

On the other hand, as a consequence of (\ref{eqxibis})
and of the independence of $H$ and $X$,
the distribution $f_\xi$ is equal to the convolution
of $f_{H}$, given by (\ref{fH}), and of $f_{X}$:
\begin{equation}
f_{\xi }(\xi )=\int_{\xi }^{1}\frac{dx}{x}f_{X}(x)
\,f_{H}\left(\frac{\xi}{x}\right)
=\frac{2\alpha}{\pi}\,\xi^{\alpha-1}\int_{\xi}^{1}dx
\frac{f_{X}(x)}{\sqrt{x^{2\alpha }-\xi ^{2\alpha }}},\label{convol}
\end{equation}
hence
\begin{equation}
\phi(\xi)=\frac{2\alpha}{\pi}\,\xi^\alpha\int_{\xi}^{1}dx
\frac{f_{X}(x)}{\sqrt{x^{2\alpha }-\xi ^{2\alpha }}}.
\label{jconvol}
\end{equation}

In summary, two conditions determine the distribution $f_X(x)$:
it obeys the symmetry property~(\ref{fx_sym}),
and the function $\phi(\xi)$, given by~(\ref{jconvol}),
obeys the symmetry property~(\ref{jphisym}).

Once the probability density $f_X$ is known,
$f_{\xi}$ is given by (\ref{convol}).
Finally, (\ref{eqMbis}), (\ref{jphidef}), and (\ref{jphisym})
imply
\begin{equation}
f_{M}(m)=\frac{1}{1-m^2}\,\phi\left(\frac{1\pm m}{2}\right).
\label{def_fM}
\end{equation}

We explore the consequences of this general setup
in the next three sections.

\section{The case $\alpha =\frac{1}{2}$}

This situation corresponds to Brownian motion.
It is easy to check that the uniform distribution on $[0,1]$,
\begin{equation}
f_X(x)=1,
\label{juni}
\end{equation}
solves the problem.
Indeed, equations~(\ref{convol}) and~(\ref{jconvol}) yield
\[
\phi(\xi)=\frac{2}{\pi}\sqrt{\xi(1-\xi)},
\qquad f_{\xi}(\xi)=\frac{2}{\pi}\sqrt{\frac{1-\xi}{\xi}},
\]
which satisfy (\ref{fxi_sym}) and (\ref{jphisym}).
Finally, by (\ref{def_fM}), the
limiting distribution of $M_{t}$ is obtained:
\begin{equation}
f_{M}(m)=\frac{1}{\pi\sqrt{1-m^2}},\label{fM}
\end{equation}
which is the arcsine law on $[-1,1]$.

All these results can be derived by more direct means,
using the fact that in the present
case the time intervals $\tau _{1},\tau _{2},\dots$ between sign changes
define a renewal process~\cite{I}.

\section{Local analysis in the persistence region}

The persistence region is defined by the condition
$M\rightarrow \pm 1$, i.e., $\xi\to0$ or $\xi\to1$.

Considering (\ref{convol}) for $\xi \rightarrow 0$ yields at once
\begin{equation}
f_{\xi}(\xi)\stackunder{\xi\rightarrow 0}{\approx}
\frac{2\alpha }{\pi}\left\langle X^{-\alpha }\right\rangle\xi^{\alpha-1},
\label{jfpers}
\end{equation}
provided the average $\left\langle X^{-\alpha }\right\rangle $
is convergent (see the comment below equation~(\ref{xalfa})).
As a consequence, using (\ref{jphidef}) and (\ref{def_fM}), we obtain
\begin{equation}
f_{M}(m)\stackunder{m\rightarrow\pm1}{\approx}C\,(1-m^2)^{\alpha-1},
\label{fMpers}
\end{equation}
with
\begin{equation}
C=\frac{2^{1-2\alpha}\,\alpha}{\pi}\left\langle X^{-\alpha }\right\rangle.
\label{jcex}
\end{equation}

The behavior of the distribution $f_X(x)$ as $x\to0$ can be determined as well.
Assuming $f_X(x)\approx A\,x^\gamma$ $(x\to0)$, and using (\ref{fx_sym}),
we obtain
\[
\phi(\xi)\stackunder{\xi\rightarrow1}{\approx}
A\,\frac{\sqrt{2\alpha}}{\pi}\int_0^{\bar{\xi}}
d\bar{x}\frac{\bar{x}^\gamma}{\sqrt{\bar{\xi}-\bar{x}}}
=A\sqrt\frac{2\alpha}{\pi}
\frac{\Gamma(\gamma+1)}{\Gamma\left(\gamma+\frac{3}{2}\right)}
\,\bar{\xi}^{\gamma+1/2},
\]
with $\bar{\xi}=1-\xi$, $\bar{x}=1-x$.
An identification with (\ref{jfpers}), using again (\ref{jphidef})
and (\ref{fxi_sym}), yields the values of $\gamma$ and $A$, hence
\begin{equation}
f_{X}(x)\stackunder{x\rightarrow0}{\approx}
\sqrt\frac{2\alpha}{\pi}
\frac{\Gamma(\alpha+1)}{\Gamma\left(\alpha+\frac{1}{2}\right)}
\left\langle X^{-\alpha}\right\rangle\,x^{\alpha-1/2}.
\label{jxpers}
\end{equation}

Let us compare the singular behavior (\ref{fMpers})
of $f_M$ in the persistence region with the beta law on $[-1,1]$ of same index:
\begin{equation}
f^{\rm beta}_M(m)=C^{\rm beta}\,(1-m^2)^{\alpha-1},
\label{jbet}
\end{equation}
where
\begin{equation}
C^{\rm beta}
=\frac{\Gamma\left(\alpha+\frac{1}{2}\right)}{\sqrt{\pi}\,\Gamma(\alpha)}.
\label{jcb}
\end{equation}
A measure of the difference between the two distributions
is provided by the enhancement factor
\begin{equation}
E=\frac{C}{C^{\rm beta}}.
\label{je}
\end{equation}

For $\alpha=\frac{1}{2}$,
the distribution $f_{M}(m)$ is the arcsine law (\ref{fM}), which is a beta law.
Equation (\ref{juni}) yields $\left\langle X^{-1/2}\right\rangle=2$,
so that $C=C^{\rm beta}=1/\pi$, and $E=1$.
The estimate (\ref{jxpers}) also agrees with (\ref{juni}).

For $\alpha\ne\frac{1}{2}$, the distribution $f_{M}(m)$ is no longer a beta
law, so that the enhancement factor $E$ is non-trivial.

\section{Asymptotic analysis for large values of $\alpha $}

\def\figun{
\vskip 8.5cm{\hskip .8cm}
\includegraphics{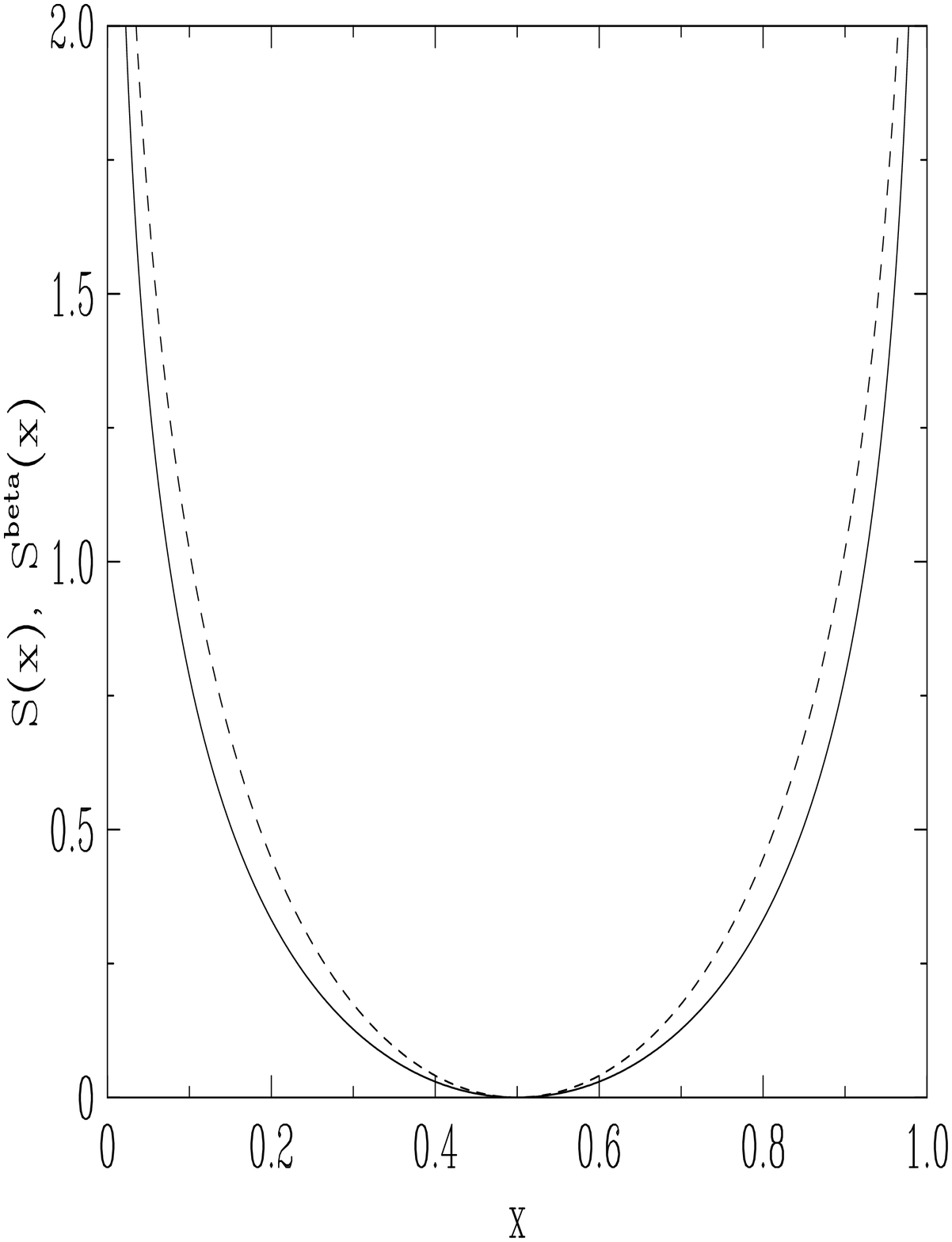}

\noindent
{\small {\bf Figure~1:}
Plot of the function $\F(x)$ characterizing the limiting distributions
of the variables $X$ and $\xi$ and of the mean magnetization $M$
in the large-$\alpha$ region, against $x$ (full line),
compared with the function $\F^{\rm beta}(x)$ associated with the beta law
(dashed line).}

\bigskip
}

\def\figde{
\vskip 8.5cm{\hskip .8cm}
\includegraphics{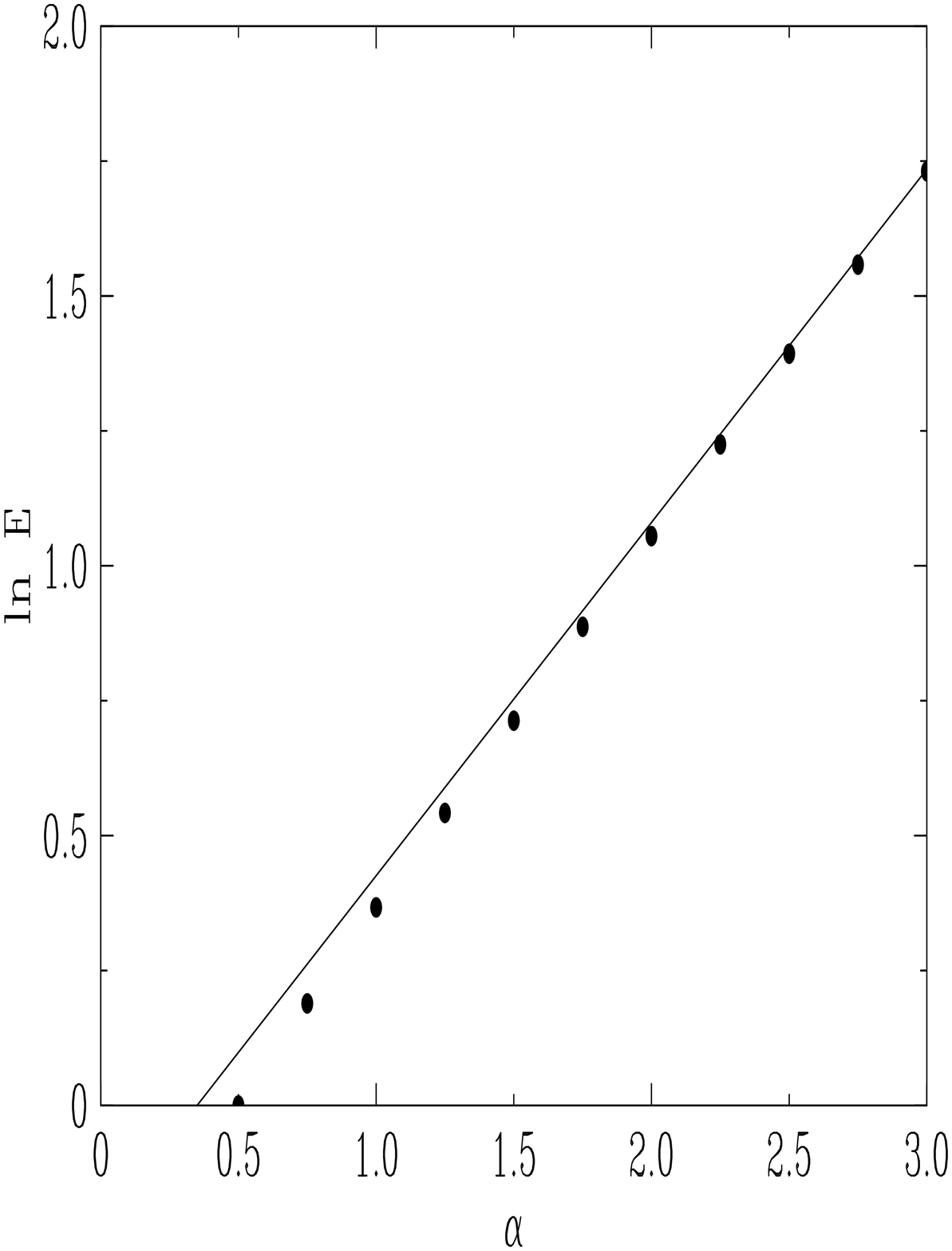}

\noindent
{\small {\bf Figure~2:}
Logarithmic plot of the enhancement factor $E$ in the persistence region,
against $\alpha$,  evaluated numerically as described in the text (symbols).
The straight line has the theoretical slope $G$ of (\ref{jgn}).}

\bigskip
}

For large values of $\alpha$,
the distributions $f_X(x)$, $f_\xi(\xi)$, and $f_M(m)$
are expected to share, at least qualitatively, some
resemblance with the beta law~(\ref{jbet}).
This observation suggests to set
\begin{equation}
f_X(x)\stackunder{\alpha\gg1}{\sim}\exp\Big(-\alpha \F(x)\Big),
\label{jax}
\end{equation}
with
\begin{equation}
\F(x)=\F(1-x),
\label{jfs}
\end{equation}
as a consequence of (\ref{fx_sym}).
The function $S(x)$ is expected to be regular, and positive,
with a minimum at $\F(\frac12)=0$, just as its counterpart
\begin{equation}
\F^{\rm beta}(x)=-\ln(4x(1-x)),
\label{jfb}
\end{equation}
associated with the beta law (\ref{jbet}).

With these hypotheses, $\phi(\xi)$, given by (\ref{jconvol}), can be estimated as follows.
Setting $x=\xi+\eps$ with $\eps\ll1$,
we have $f_X(x)\approx\e^{-\alpha \F(\xi)-\alpha\eps \F'(\xi)}$
and $x^{2\alpha}-\xi^{2\alpha}\approx\xi^{2\alpha}(\e^{2\alpha\eps/\xi}-1)$.
The change of integration variable from $x$ to $z=2\alpha\eps/\xi$ yields
\begin{equation}
\phi(\xi)\stackunder{\alpha\gg1}{\approx}P(\xi)f_X(\xi),
\label{jaxi}
\end{equation}
with
\[
P(\xi)=\frac{\xi}{\pi}
\int_0^\infty dz \frad{\e^{-\xi \F'(\xi) z/2}}{\sqrt{\e^z-1}}.
\]
Setting $y=\e^{-z}$, and returning to the variable $x$, we finally obtain
\begin{equation}
P(x)=\frad{x}{\sqrt{\pi}}
\frad{\Gamma\left(\frad{1}{2}+\frad{x \F'(x)}{2}\right)}
{\Gamma\left(1+\frad{x \F'(x)}{2}\right)},
\label{jpex}
\end{equation}
provided the arguments of both Gamma functions are positive
(this will indeed be the case).
Note that (\ref{jpex}) does not involve the parameter $\alpha$ any more.

Omitting again pre-exponential factors, (\ref{jaxi}) implies
\begin{equation}
f_M(m)\stackunder{\alpha\gg1}{\sim}
\exp\left(-\alpha\F\left(\frad{1\pm m}{2}\right)\right).
\label{jam}
\end{equation}

In the regime of large $\alpha$,
the three distributions of interest are therefore given by a single
function $\F(x)$.
The problem then amounts to finding $\F(x)$,
with the symmetry property (\ref{jfs}),
and such that the corresponding function $P(x)$,
given by (\ref{jpex}), obey
\begin{equation}
P(x)=P(1-x),
\label{jpsym}
\end{equation}
as a consequence of (\ref{jphisym}).
The function $\F(x)$ is entirely determined by the above conditions.
This property is more evident in the present regime than in the general case
of section~\ref{jsecgen}, because (\ref{jpex}) is explicit,
while (\ref{jconvol}) is an integral relationship.

Let us first investigate
the behavior of $\F(x)$ for $x\to\frac12$, i.e., $m\to0$,
corresponding to the center of the distributions.
Inserting the expansion
\[
\F(x)=c_2\left(x-\frac12\right)^2+c_4\left(x-\frac12\right)^4+\cdots
\]
in (\ref{jpex}), (\ref{jpsym}),
and expanding the Gamma functions accordingly, we obtain
\begin{equation}
c_2=\frac{2}{\ln2},\qquad
c_4=\frac{4}{3\ln2}+\frac{\pi^2}{3(\ln2)^3}-\frac{\zeta(3)}{(\ln2)^4},
\qquad\dots
\label{jc}
\end{equation}
and
\[
P(x)=\frac12+\left(\frac{\pi^2}{12(\ln2)^2}-3\right)\left(x-\frac12\right)^2
+\cdots
\]

To leading order, keeping only the quadratic term in $\F(x)$,
we find that the bulk of the distributions is asymptotically
given by narrow Gaussians for $\alpha$ large, namely
\[
f_X(x)\stackunder{\alpha\gg1}{\sim}
f_\xi(x)\stackunder{\alpha\gg1}{\sim}
\exp\left(-\frac{2\alpha}{\ln2}\left(x-\frac12\right)^2\right),\qquad
f_M(m)\stackunder{\alpha\gg1}{\sim}
\exp\left(-\frac{\alpha}{2\ln2}m^2\right).
\]
The latter result is in agreement with the expressions (\ref{rec_mkbis})
of the moments of $M$,
which behave as $\left\langle M^2\right\rangle\approx(\ln2)/\alpha$,
$\left\langle M^4\right\rangle\approx3(\ln2)^2/\alpha^2$, and so on,
for $\alpha\gg1$.

It is also worthwhile noticing that the beta law (\ref{jbet}), (\ref{jfb})
also becomes a narrow Gaussian for $\alpha$ large.
We have $\F^{\rm beta}(x)\approx4(x-\frac12)^2$
and $f^{\rm beta}_M(m)\approx\e^{-\alpha m^2}$,
so that the beta law misses a finite factor $2\ln2\approx1.3862$
in the variance of the mean magnetization.

The expression (\ref{jc}) of the subleading amplitude $c_4$,
involving Riemann's zeta function,
shows however that the function $\F(x)$ is altogether non-trivial.

\figun

Let us now turn to the behavior of $\F(x)$ deep in the tails
of the distributions, i.e., for $x\to0$ or 1,
or $m\to\pm1$, corresponding to the persistence region.
The general result (\ref{fMpers}) shows that $\F(x)\approx -\ln x$
has a logarithmic divergence as $x\to0$.
As a consequence, the Gamma function in the numerator of expression
(\ref{jpex}) for $P(x)$ becomes singular, as its argument goes to zero.
Furthermore, in the same expression for $P(1-x)$,
the arguments of both Gamma functions tend to infinity.
A careful treatment of (\ref{jpex})
yields the more complete expansions as $x\to0$
\begin{equation}
\F(x)=-\ln x+\F_0+2\sqrt{\frac{2x}{\pi}}+\cdots,\qquad
P(x)=\sqrt{\frac{2x}{\pi}}+\cdots,
\label{jff}
\end{equation}
while the constant $\F_0$ cannot be predicted by this local analysis.
The square-root behavior of $P(x)$ and its prefactor agree 
with the general results (\ref{jfpers}), (\ref{jxpers}).

We have numerically determined the solution of (\ref{jpex}), (\ref{jpsym})
over the whole range $0<x<\frac12$,
obtaining thus accurate values of $\F(x)$.
This approach yields in particular $\F_0\approx-2.0410$.
Figure~1 shows a plot of the function $\F(x)$ thus obtained,
compared with $\F^{\rm beta}(x)$.

As the amplitude $C^{\rm beta}$
of the beta law (\ref{jbet}) remains of order unity,
within exponential accuracy,
the result (\ref{jff}) for $\F(x)$ implies that
the amplitude $C$ of the power law (\ref{fMpers}) in the persistence region,
and the enhancement factor $E$ defined in (\ref{je}),
blow up exponentially, as
\begin{equation}
C\stackunder{\alpha\gg1}{\sim}
E\stackunder{\alpha\gg1}{\sim}\exp(G\alpha),
\label{jg}
\end{equation}
with
\begin{equation}
G=\lim_{x\to0}(\F^{\rm beta}(x)-\F(x))
=-\F_0-2\ln2\approx0.6547.
\label{jgn}
\end{equation}

In order to test the relevance of this large-$\alpha$ approach,
we have evaluated numerically $E$
for various values of the parameter $\alpha$,
and compared the results with the exponential law (\ref{jg})
predicted for large $\alpha$.
The computation of $E$ can be done
in (at least) two different ways.

The first method consists in directly evaluating the limit
\[
E=\lim_{n\to\infty}\frac{\left\langle M^{2n}\right\rangle}
{\left\langle M^{2n}\right\rangle^{\rm beta}}.
\]
The moments of the beta law (\ref{jbet}) read
\[
\left\langle M^{2n}\right\rangle^{\rm beta}
=\frac{\Gamma\left(\alpha+\frac12\right)\Gamma\left(n+\frac12\right)}
{\sqrt{\pi}\,\Gamma\left(n+\alpha+\frac12\right)},
\]
while the true moments $\left\langle M^{2n}\right\rangle$
are determined from (\ref{recur1bis}), (\ref{recur2bis}), and (\ref{jme}),
up to some maximal order, typically $n\max=100$ to 150,
beyond which numerical accuracy rapidly deteriorates,
because the computation of $\left\langle M^{2n}\right\rangle$
involves alternating sums.

The second method consists in combining (\ref{jcex}) and (\ref{jcb}),
getting thus
\[
E=\frac{2^{1-2\alpha}}{\sqrt{\pi}}\,
\frac{\Gamma(\alpha+1)}{\Gamma\left(\alpha+\frac12\right)}
\left\langle X^{-\alpha} \right\rangle,
\]
and in evaluating $\left\langle X^{-\alpha} \right\rangle$ as
\begin{equation}
\left\langle X^{-\alpha} \right\rangle
=\left\langle (1-X)^{-\alpha} \right\rangle
=\frac{1}{\Gamma(\alpha)}
\sum_{n=0}^\infty\frac{\Gamma(n+\alpha)}{n!}\,x_n.
\label{xalfa}
\end{equation}
The behavior (\ref{jxpers}) implies that the $x_n$ decay as $n^{-\alpha-1/2}$,
so that the term of order $n$ in the above sum decays as $n^{-3/2}$.
Hence this sum is convergent, and truncating it at some order $n\max$
brings a correction proportional to $n\max^{-1/2}$.
The $x_n$ are again determined from (\ref{recur1bis}) and (\ref{recur2bis}).
A linear extrapolation in $n\max^{-1/2}$ of the results of both schemes
turns out to yield consistent results.
We have for instance $E\approx1.443$ for $\alpha=1$.

\figde

Figure~2 shows our numerical results for the enhancement factor $E$,
for values of $\alpha$ up to 3.
The comparison with the exponential law (\ref{jg}) is convincing,
in spite of the moderate values of $\alpha$ used.

\section{Revisiting the work of Dhar and Majumdar}

\label{revisit}

\subsection{Using the method of section \ref{reminder}}

Let us first show that the expressions (\ref{rec_mkbis})
for the even moments of $M$ are the
same as those obtained by using (\ref{rec_mk}), with the
expression of $\hat{A}(s)$ appropriate to the process (\ref{process}),
as done in \cite{maj}.

The autocorrelation $A(|\Delta T|)=\left\langle \sigma _{T}\sigma
_{T+\Delta T}\right\rangle $ of the sign process $\sigma_t$
in the logarithmic scale of time $T=\ln t$ reads
$A(T)=(2/\pi)\arcsin(\text{e}^{-\alpha \mid T\mid })$ \cite{maj},
with Laplace transform
\begin{equation}
\hat{A}(s)=\frac{1}{s}\left[ 1-\frac{1}{\pi }B\left( \frac{s}{2\alpha }+%
\frac{1}{2},\frac{1}{2}\right) \right].
\label{Ahat}
\end{equation}
We notice that $\hat{A}(s)$ is related to $h(s,\alpha )$, defined in (\ref
{hsa}), by
\begin{equation}
h(s,\alpha )=1-s\hat{A}(s),
\label{relation}
\end{equation}
where the right side is equal to $2g(s)/s$ by (\ref{gs}).
Using this identity, it is easy to check that the
moments (\ref{rec_mk}) obtained by the method of section \ref{reminder},
with $\hat A_k$ given by~(\ref{Ahat}) for $s=k$ integer,
are identical to the moments (\ref{rec_mkbis}) obtained by the method of
the present work.

It is, however, not possible to identify the intermediate results of both
methods, as can be seen by comparing respectively equation (\ref{fkt}) to
equations (\ref{def_fk}) and (\ref{fls}), and equation (\ref{hsa}) to
equation (\ref{flambdas}).
This demonstrates the formal character of the
application of the method of section \ref{reminder} to the process (\ref
{process}).
(See also the discussion in section 3.3.)

\subsection{Comments on the results obtained using Kac's formalism}

A first comment is that the recursion relations for the coefficients $c_{k}$
appearing in equations
(14) of ref. \cite{maj} can be easily recognized to be identical to the
recursion relations (\ref{recur1bis}), (\ref{recur2bis}) for the $x_{k}$,
by noting the correspondences
\begin{eqnarray}
c_{k} &=&\frac{2^{k}}{k!D_{-k/\alpha }(0)}x_{k}, \nonumber \\
\frac{D_{-k/\alpha +1}(0)}{D_{-k/\alpha }(0)} &=&\frac{\sqrt{2\pi }}{\alpha }%
\frac{k\,h_{k}^{(\alpha )}}{2}. \label{identif}
\end{eqnarray}

The second comment concerns the continuity conditions expressed in
equations (13) of ref. \cite{maj}.
Using (\ref{identif}), these conditions yield, with the notations of the
present work,
\begin{eqnarray}\nonumber
\left\langle \text{e}^{a(2X-1)}\right\rangle &=&\left\langle \text{e}%
^{a(1-2X)}\right\rangle,
\\
\nonumber
\left\langle \xi \text{e}^{a(2\xi -1)}\right\rangle &=&\left\langle \xi
\text{e}^{a(1-2\xi )}\right\rangle.
\end{eqnarray}
These equations hold for $a$ arbitrary, hence they are equivalent to
(\ref{fx_sym}) and (\ref{fxi_sym}), respectively.

\section{Summary and discussion}

In this paper we have revisited and extended
the work of Dhar and Majumdar \cite{maj}.
Besides providing a new recursive determination of
the moments of the mean magnetization $M$,
the present study leads to a functional integral equation for
the distribution of the latter quantity.
This framework allows a local analysis of this distribution,
and of other relevant quantities,
in the persistence region $(M_t\to\pm1)$,
as well as a detailed investigation of the regime where $\alpha$ is large.

The present work casts new light 
on the status of the expressions (\ref{rec_mk}) for the moments of $M$.
The method recalled in section~\ref{reminder}, which leads to these relations,
can be applied to any smooth process
for which the intervals of time $\l_N$ between sign changes are independent,
on a logarithmic scale.
For this class of processes $\langle\l_N\rangle=\bar{\l}$ is finite (i.e., non-zero),
and the mean number of sign changes between $0$ and $t$ scales as
$\left\langle N_t \right\rangle\approx(\ln t)/\bar{\l}$.

Relations (\ref{rec_mk}) are also verified
for the class of processes considered in this work.
This was observed in \cite{maj} (by comparing the expressions thus found
to those obtained by another method,
based on a formalism due to Kac), and justified
by the absence of $\,\bar{\l}\,$ in equations (\ref{rec_mk}).
Yet, as discussed in section 3.3, 
in the present case there is no obvious reason
to work with a logarithmic scale of time, since $\langle\l_N\rangle$
asymptotically vanishes, and the mean number of sign changes scales as
$\left\langle N_t \right\rangle\approx2\,\pi^{-1/2}\,t^\alpha$ \cite{I}.
(See also the discussion in section 10.1.)

Most of the effort of the present work was to provide
a new derivation of (\ref{rec_mk}) for the class of processes (\ref{process}).
Our approach relies on the fact that the time intervals $\tau'_n$
between two sign changes of the process (\ref{process}) form a renewal process
(the $\tau'_n$ are independent, identically distributed random variables).
The derivation proceeds in two steps.
First, relations (\ref{rec_mkbis})
for the $\langle M^k\rangle$ are obtained;
then, using (\ref{relation}), equations (\ref{rec_mkbis}) yield (\ref{rec_mk}).
This extends the range of applicability of relations (\ref{rec_mk}).
Note that for diffusion (in the independent-interval approximation) 
(see section 3.2), relations (\ref{rec_mk}) hold
but neither (\ref{rec_mkbis}) nor (\ref{relation}) do.

We conclude by a few additional comments.

In passing, let us mention another equivalent formulation of (\ref{relation}),
namely that the two-time autocorrelation of the sign process reads,
with $t<t'$,
\begin{equation}
C(t,t')=\int_0^{t/t'}dx\,f_H(x).
\label{CH}
\end{equation}

Another situation where (\ref{rec_mk}), (\ref{rec_mkbis}),
and (\ref{relation}) or (\ref{CH}) hold is for the renewal processes
considered in \cite{I} (provided $\theta <1$),
which are yet another deformation of Brownian motion.

Note that the first relation of (\ref{rec_mk}),
$\left\langle M^{2}\right\rangle =\hat{A}_{1}$,
holds whenever the two-time autocorrelation function
is a scaling function of the ratio of the two times \cite{dg},
while the first relation of (\ref{rec_mkbis}),
$\left\langle M^2\right\rangle=1-\left\langle H\right\rangle$,
does not in general.
For instance, for the random acceleration problem,
using results of ref. \cite{epl},
we find $\left\langle M^2\right\rangle=3\sqrt3/\pi-1\approx0.653986$
and $1-\left\langle H\right\rangle\approx 0.791335$.

This work also underlines
the importance of the random variables $X$ and $H$.
The distribution of the latter is known exactly in the present case.
This quantity, which is a natural one to consider for 
Brownian motion~\cite{feller},
and more generally for renewal processes \cite{I},
appears also in the context of phase ordering~\cite{red}.

As mentioned in the introduction,
the process (\ref{process}) has been proposed \cite{sebastian}
as a Markovian approximation to fractional Brownian motion.
Let us compare the expressions of $\langle M^2\rangle$ for these
two processes.
For the present model we have (see~(\ref{rec_mkbis}))
\begin{equation}
\langle M^2\rangle=1-\frac{1}{\sqrt{\pi}}
\frac{\Gamma\left(\frad{1}{2\alpha}+\frad{1}{2}\right)}
{\Gamma\left(\frad{1}{2\alpha}+1\right)},
\label{1j}
\end{equation}
while for fractional Brownian motion, with H\"older index $0<h<1$, we have
\begin{equation}
\langle M^2\rangle=\frac{2}{\pi}\int_0^1
dx\,\arcsin\frac{x^{2h}+1-(1-x)^{2h}}{2x^h}.
\label{2j}
\end{equation}
The correspondence between the two processes is made by identifying their
persistence exponents: $\theta=\alpha=1-h$.
For $\theta=1/2$, we have $\langle M^2\rangle=1/2$ in both cases.
For $\theta=1$, (\ref{1j}) yields $\langle M^2\rangle=1-2/\pi\approx0.363380$,
while (\ref{2j}) yields $\langle M^2\rangle=1/3$.
For $\theta\to0$, we have $\langle M^2\rangle=1-c\sqrt\theta$,
with (\ref{1j}) yielding $c=\sqrt{2/\pi}\approx0.797885$,
and (\ref{2j}) yielding $c\approx0.812233$.
The distributions of the mean magnetization for the two processes
are therefore expected to be rather similar (for $0<\theta<1$).

Finally, let us comment on the changes in behaviour induced by letting 
the persistence exponent $\alpha$ vary,
and compare the present process to other ones in this respect.
The distribution of $M$ shows a change in shape as $\alpha$ increases, 
the most probable value of the mean magnetization shifting 
from the edges to the center \cite{maj}.
More precisely, as shown in section 8, as long as $\alpha<1$, 
$f_M(m)$ diverges at $m\to\pm1$, while for $\alpha>1$ it 
vanishes at these points (see equation~(\ref{fMpers})).
However for any arbitrary value of $\alpha$ the magnetization $M$
remains distributed.

This behaviour is actually generic, whenever the two-time autocorrelation
function of the process is asymptotically
a function of the ratio of the two time variables \cite{dg}.
In particular, this is so for diffusion.
In the independent-interval approximation the persistence exponent 
$\theta(D)\approx 0.1454\sqrt{D}$
increases without bounds when the dimension of space $D$ is large 
\cite{majdiffu,derrdiffu}.
As originally noted in \cite{dg}, 
as long as $\theta<1$ the density $f_M(m)$ diverges 
at the edges, while it vanishes there if $\theta>1$.
This was also emphasized  in \cite{newman}, on the basis of scaling
arguments, and recently confirmed by direct numerical computations \cite{newman3}.

In contrast, there are other processes 
for which the change in behaviour at $\theta=1$ is more radical.
For fractional Brownian motion, $\theta=1$ appears as a maximum persistence
exponent.
For the renewal processes considered in ref.~\cite{I}, 
the mean magnetization possesses a non-trivial asymptotic 
distribution only if $\theta<1$.

\newpage
\appendix

\section{Properties of the binomial operator $\mathcal{B}$}

The aim of this Appendix is to prove the following property, used in section
5.2.
Assume that the sequence $x_{k}$ satisfies
\begin{equation}
x_{k}=\mathcal{B}(f_{k}x_{k}),
\label{Bfx}
\end{equation}
with $x_{0}=f_{0}=1$, and where
$\B(x_k)=\sum_{j=0}^{k}\binom{k}{j}\left( -1\right)^{j}\,x_{j}$
(see (\ref{jb})).
Then
\begin{eqnarray}
x_{k}(1+f_{k})&=&\mathcal{B}(x_{k}(1+f_{k}))\qquad {\hskip 10pt}(k\text{ odd}),
\label{odd2}
\\
x_{k}(1-f_{k})&=&-\mathcal{B}(x_{k}(1-f_{k}))\qquad(k\text{ even}),
\label{even2}
\end{eqnarray}
which are respectively equations (\ref{odd})
and
(\ref{even}) in the text.

\subsection{Basic properties}

In order to prove (\ref{odd2}) and (\ref{even2})
we need the following auxiliary properties.

First, $\B$ is its own inverse:
\begin{equation}
\B=\B^{-1}.
\label{jainv}
\end{equation}
A combinatorial proof of this result can be found in ref. \cite{knuth}.
An alternative proof is obtained by noting
that the action of $\B$ on exponential sequences $x_k=y^k$ reads
\begin{equation}
\B(y^k)=(1-y)^k.
\label{jae}
\end{equation}
This relation is invariant in the change of $y$ to $1-y$,
hence (\ref{jainv}) follows.

Then, we have the properties
\begin{eqnarray}
&&x_k=\B(x_k) \quad\text{for }k\text{ even}
{\hskip 23.25pt}\text{implies} \quad x_k=\B(x_k)\quad\text{for all }k,
\label{j1}\\
&& x_k=\B(x_k) \quad\text{for }k\text{ odd}
{\hskip 26.5pt}\text{implies} \quad x_k=\B(x_k)\quad\text{for all }k,
\label{j2}\\
&& x_k=-\B(x_k) \quad\text{for }k\text{ even}
{\hskip 14pt}\text{implies} \quad x_k=-\B(x_k)\quad\text{for all }k,
\label{j3}\\
&& x_k=-\B(x_k) \quad\text{for }k\text{ odd}
{\hskip 17.25pt}\text{implies} \quad x_k=-\B(x_k)\quad\text{for all }k.
\label{j4}
\end{eqnarray}

Before giving the proofs, let us explicit the meaning of (\ref{j1}-\ref{j4}).

Let us take the example of (\ref{j1}).
By hypothesis, the sequence $x_k$ satisfies the condition
$x_k=\B(x_k)$ for $k$ even, with $x_{0}$ arbitrary,
which is equivalent to saying that
\begin{equation}
\sum_{j=0}^{k-1}\binom{k}{j}\left( -1\right) ^{j}\,x_{j}=0
\qquad(k\text{ even}).
\label{rec1}
\end{equation}
This recursion determines $x_{k}$ for $k$ odd in terms of the $x_{\ell}$
with $\ell=0,\dots,k-1$ even:
\[
x_1=\frac12x_0,\qquad x_3=\frac32x_2-\frac14x_0,\qquad
x_5=\frac52x_4-\frac52x_2+\frac12x_0,\qquad\dots
\]
The property (\ref{j1}) states that
$x_k=\B(x_k)$ for $k$ odd,
or equivalently,
\[
2x_k=\sum_{j=0}^{k-1}\binom{k}{j}\left( -1\right) ^{j}\,x_{j}
\qquad(k\text{ odd}),
\]
which provides an infinite number of consistency relations
amongst the $x_{k}$ satisfying~(\ref{rec1}).

Similarly, taking the example of (\ref{j3}), by hypothesis we have
$x_k=-\B(x_k)$ for $k$ even, with $x_0=0$, which is is equivalent to
\begin{equation}
2x_k=-\sum_{j=0}^{k-1}\binom{k}{j}\left( -1\right) ^{j}\,x_{j}
\qquad(k\text{ even}).
\label{rec2}
\end{equation}
This recursion determines $x_{k}$ for $k$ even in terms of the $x_{\ell}$
with $\ell=1,\dots,k-1$ odd:
\[
x_2=x_1,\qquad x_4=2x_3-x_1,\qquad x_6=3x_5-5x_3+3x_1,\qquad\dots
\]
The property (\ref{j3}) states that
$x_k=-\B(x_k)$ for $k$ odd,
or equivalently,
\[
\sum_{j=0}^{k-1}\binom{k}{j}\left( -1\right)^{j}\,x_{j}=0
\qquad(k\text{ odd}),
\]
which provides an infinite number of consistency relations
amongst the $x_{k}$ satisfying~(\ref{rec2}).

We now prove the properties (\ref{j1}-\ref{j4}).
In order to do so, let us define,
for a given sequence $x_k$, the Laurent series
\[
F(z)=\sum_{k=0}^\infty x_k\,z^{-k},\qquad
G(z)=\sum_{k=0}^\infty \B(x_k)\,z^{-k}.
\]
We assume that these series are
convergent for $\vert z\vert$ larger than some radius $R$.
This happens e.g. if the $x_k$ are bounded.

The functions $F(z)$ and $G(z)$ are related to each other by
\begin{eqnarray}
F(z)=\frac{z}{z-1}\,G(1-z),\label{ja1}\\
G(z)=\frac{z}{z-1}\,F(1-z),\label{ja2}
\end{eqnarray}
as we now show.
We have
\[
x_k=\oint\frac{dy}{2\pi iy}\,y^k\,F(y),
\]
hence, using (\ref{jae}),
\[
\B(x_k)=\oint\frac{dy}{2\pi iy}\,(1-y)^k\,F(y),
\]
so that
\[
G(z)=\oint\frac{dy}{2\pi iy}\,\frac{z}{y+z-1}\,F(y).
\]
This integral is equal to the contribution of the pole at $y=1-z$,
yielding (\ref{ja2}), from which
(\ref{ja1}) follows.
The symmetric form of the formulas (\ref{ja1}), (\ref{ja2})
is due to the property (\ref{jainv}).

\medskip
\noindent{\bf Proof of (\ref{j1}) and (\ref{j2})}

The hypothesis in (\ref{j1}) implies $F(z)+F(-z)=G(z)+G(-z)$, i.e.,
\begin{equation}
\Phi(z)=-\Phi(-z),
\label{jphi1}
\end{equation}
with, using (\ref{ja2}),
\[
\Phi(z)=F(z)-G(z)=F(z)-\frac{z}{z-1}F(1-z).
\]
Therefore
$(z-1)\Phi(z)+z\Phi(1-z)=0$,
which can be rewritten, using (\ref{jphi1}), as
$$
\frac{\Phi(z)}{z}=\frac{\Phi(z-1)}{z-1}.
$$
The function $\Phi(z)/z$ is thus periodic, with unit period,
and decaying at infinity, as we have $\Phi(z)/z\approx(x_0-2x_1)/z^2$ a priori.
We conclude that $\Phi(z)=0$ identically, that is $F(z)=G(z)$,
implying the property (\ref{j1}).

For the case where the $x_{k}=\left\langle X^{k}\right\rangle$
are the moments of a random variable $X$, with density $f_{X}$ on $[0,1]$,
an alternative proof of (\ref{j1}) is as follows.
The hypothesis in (\ref{j1}) expresses the property
\[
\left\langle X^{k}\right\rangle =\left\langle(1-X)^{k}\right\rangle
\qquad(k\text{ even}).
\]
As both random variables $X$ and $1-X$ are positive,
this last condition is sufficient to imply $f_{X}(x)=f_{X}(1-x)$, hence
$\left\langle X^{k}\right\rangle =\left\langle(1-X)^{k}\right\rangle$
for all $k$, which proves (\ref{j1}).

The proof of the second property, (\ref{j2}), is very similar.
The hypothesis in (\ref{j2}) implies
\[
\frac{\Phi(z)}{z}=-\frac{\Phi(z-1)}{z-1}.
\]
The function $\Phi(z)/z$ is therefore periodic, with period two,
and decaying at infinity, hence $\Phi(z)=0$ identically.

\medskip
\noindent{\bf Proof of (\ref{j3}) and (\ref{j4})}

The hypothesis in (\ref{j3}) implies $F(z)+F(-z)=-(G(z)+G(-z))$, i.e.,
\begin{equation}
\Psi(z)=-\Psi(-z),
\label{jpsi1}
\end{equation}
with, using (\ref{ja2}),
\[
\Psi(z)=F(z)+G(z)=F(z)+\frac{z}{z-1}F(1-z).
\]
Therefore
$(z-1)\Psi(z)-z\Psi(1-z)=0$,
which can be rewritten, using
(\ref{jpsi1}), as
\[
\frac{\Psi(z)}{z}=-\frac{\Psi(z-1)}{z-1}.
\]
The function $\Psi(z)/z$ is thus again periodic, with period two,
and decaying at infinity, hence identically zero.
The proof of the fourth property, (\ref{j4}), is very similar.

\subsection{Proofs of equations (\ref{odd2}) and (\ref{even2})}

Equation (\ref{Bfx}) implies the relations
\begin{eqnarray}
x_{k}(1+f_{k}) &=&\sum_{j=0}^{k-1}\binom{k}{j}\left( -1\right)
^{j}\,f_{j}\,x_{j}\qquad(k\text{ odd)},\label{direct1} \\
x_{k}(1-f_{k}) &=&\sum_{j=0}^{k-1}\binom{k}{j}\left( -1\right)
^{j}\,f_{j}\,x_{j}\qquad(k\text{ even)},
\label{direct2}
\end{eqnarray}
which determines the $x_k$ recursively.
We have thus
\begin{equation}
x_{1}=\frac{1}{1+f_{1}},\qquad
x_{2}=\frac{1-f_{1}}{(1+f_{1})(1-f_{2})},\qquad\dots
\label{rec_xk}
\end{equation}
Since the operator $\mathcal{B}$ is its own inverse,
(\ref{Bfx}) is equivalent to
$f_{k}\,x_{k}=\mathcal{B}(x_{k})$,
which itself implies
\begin{eqnarray}
x_{k}(1+f_{k}) &=&\sum_{j=0}^{k-1}\binom{k}{j}\left( -1\right)
^{j}\,x_{j}\qquad(k\text{ odd)},\label{invers1} \\
x_{k}(f_{k}-1) &=&\sum_{j=0}^{k-1}\binom{k}{j}\left( -1\right)
^{j}\,x_{j}\qquad(k\text{ even)}.\label{invers2}
\end{eqnarray}
Comparing (\ref{direct1}) and (\ref{invers1}) shows that
\[
\sum_{j=0}^{k-1}\binom{k}{j}\left( -1\right) ^{j}\,x_{j}(1-f_{j})=0\qquad
(k\text{ odd}),
\]
hence, using the property (\ref{j4}),
\[
x_{k}(1-f_{k})=-\mathcal{B}(x_{k}(1-f_{k}))\qquad(k\text{ even}),
\]
which is equation (\ref{even2}). Similarly, comparing (\ref
{direct2}) and (\ref{invers2}) shows that
\[
\sum_{j=0}^{k-1}\binom{k}{j}\left( -1\right) ^{j}\,x_{j}(1+f_{j})=0\qquad
(k\text{ even)},
\]
or, using the property (\ref{j1}),
\[
x_{k}(1+f_{k})=\mathcal{B}(x_{k}(1+f_{k}))\qquad(k\text{ odd}),
\]
which is equation (\ref{odd2}).

\end{document}